\documentclass[a4paper,pdftex]{article}

\usepackage[utf8]{inputenc} 
\usepackage[T1]{fontenc}    
\usepackage[colorlinks=true,linkcolor=black,anchorcolor=black,citecolor=black,filecolor=black,menucolor=black,runcolor=black,urlcolor=black]{hyperref}       
\usepackage{url}            
\usepackage{graphicx}       
\graphicspath{{media/}}     

\usepackage{a4wide}       

\begin{document}
\title{Towards the optimization of ballistics in proton therapy using genetic algorithms: implementation issues}
\author{F~Smekens$^{\dag}$, N~Freud$^{\dag}$, B~Sixou$^{\dag}$, G~Beslon$^{\href{https://orcid.org/0000-0001-8562-0732}{\includegraphics[scale=0.06]{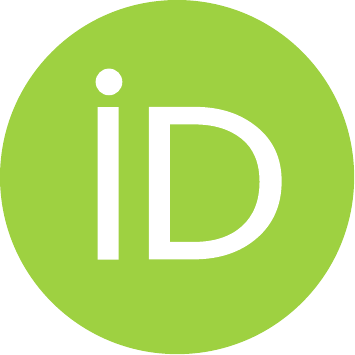}}\ddag}$, \& JM~Létang$^{\href{https://orcid.org/0000-0003-2583-782X}{\includegraphics[scale=0.06]{orcid.pdf}}\dag}$\\
  {\small $^\dag$ CREATIS, CNRS UMR5220, Inserm U1294, INSA Lyon,}\\
  {\small Universit\'e Claude Bernard Lyon 1, UJM Saint Étienne, F-69373 Lyon, France}\\
  {\small $^\ddag$ LIRIS, CNRS UMR5205, INSA-Lyon, Université Claude Bernard Lyon 1,}\\
  {\small Université Lumière Lyon 2, École Centrale de Lyon, F-69621 Villeurbanne, France}\\
}
\date{{\small Version typeset: \today}}
\maketitle

\begin{abstract}
Proton therapy is a treatment modality in fast development. The dose delivered to the planning target volume by proton beams is highly conformal, sparing organs at risk and normal tissues. New treatment planning systems adapted to spot scanning techniques have been recently proposed to simultaneously optimize several fields and thus improve dose delivery.
In most cases the only parameters considered in the optimization are the field fluences whereas other parameters -- e.g. beam directions -- are based on the experience of medical physicists and are limited by technological constraints. It is likely that future irradiation facilities will offer improved treatments as well as increased complexity in the planning.

In this paper, we investigate a new optimization framework based on a genetic algorithm approach. This tool is intended to make it possible to explore new schemes of treatment delivery, possibly with future enhanced technologies. The optimization framework is designed to be versatile and to account for many degrees of freedom, without any {\it a priori} technological constraint. To test the behavior of our algorithm, we propose in this paper, as an example, to optimize beam fluences, target points and irradiation directions at the same time.

The proposed optimization routine takes typically into account several thousands of spots of fixed size. The evolution is carried out by the three standard genetic operators: mutation, crossover and selection. The figure-of-merit (or fitness) is based on an objective function relative to the dose prescription to the tumor and to the limits set for organs at risk and normal tissues. Fluence optimization is carried out via a specific scheme based on a plain gradient with analytical solution. Several specific genetic algorithm issues are addressed: (i) the mutation rate is tuned to balance the search and selection forces, (ii) the initial population is selected using a bootstrap technique and (iii) to scale down the computation time, dose calculations are carried out with a fast analytical ray tracing method and are multi-threaded.

In this paper implementation issues of the optimization framework are thoroughly described. The behavior of the proposed genetic algorithm is illustrated in both elementary and clinically-realistic test cases.
\end{abstract}

\hrulefill

\smallskip
\centerline{\textsl{Correspondence should be addressed to: \texttt{jean.letang@creatis.insa-lyon.fr}}}

\hrulefill

\clearpage
\section{Introduction}

Proton therapy is a fast evolving treatment modality. Proton beams deliver a dose highly conformal to the prescribed target volume, sparing organs at risk and normal tissues. Recently, new treatment planning systems (TPS) adapted to the spot scanning technique have been proposed to simultaneously optimize several fields and improve the quality of treatment~\cite{Gemmel2008}. A typical treatment fraction generally consists of several steps including computed tomography (CT) acquisition, contouring of the tumor and critical structures, inverse planning and treatment delivery.

In this paper, we focus our study on inverse planning and propose an optimization technique based on a genetic algorithm (GA), which makes it possible to explore new schemes of treatment delivery, possibly with future enhanced technologies, without any {\it a priori} technological constraint. This approach is designed to be versatile and to account for many degrees of freedom (beam fluences, target points and irradiation directions). Our inverse planning scheme is divided into two parts: the forward dose calculation engine (section~\ref{sect_dose_simulation}) and the optimization algorithm itself (section~\ref{sect_genetic_algorithm}). The performance of the proposed inverse planning method is analyzed in a few test cases in section~\ref{sect_results}.

As an introduction, the state-of-the-art planning techniques used in the case of active treatment delivery (spot scanning) are first introduced in subsection~\ref{ssec:sspt}, then some of the trails offered by intensity modulated radiotherapy (IMRT) are summarized in subsection~\ref{ssec:imrt}. Finally, the scheme of the proposed optimization routine based on a genetic algorithm is detailed in subsection~\ref{ssec:hga}.

\subsection{Spot scanning planning techniques}
\label{ssec:sspt}

In the spot scanning active delivery mode (intensity modulated particle therapy or IMPT), a proton beam with a transverse Gaussian profile (typically 5--10~mm FWHM) is scanned in two dimensions using magnetic devices. The proton penetration depth in matter is adjusted by tuning the beam energy.

The current planning techniques~\cite{Lomax1999,Lomax2004,Bourhaleb2008} are based on analytical algorithms which consist in automatically filling the target volume with numerous spots from the distal edge to the proximal side of the tumor. If necessary, a conjugate gradient is used to fine-tune beam fluences.
New algorithms (still based on analytical techniques) making it possible to optimize several fields simultaneously have been proposed recently, leading to increased sparing of healthy tissues~\cite{Gemmel2008}.
It is worthy of note that in all previously-cited works, the only parameters optimized are the beam fluence values, whereas all other parameters (beam source position, spot size and spacing) are selected manually by the medical physicist.

\subsection{IMRT planning techniques}
\label{ssec:imrt}
Optimization techniques in IMRT constitute a very useful basis for our own work.
In the optimization process, we distinguish between the algorithm itself and the fitness evaluation, the aim of which is to assess the beam layout with respect to the optimization objectives.
It is often proposed to adapt the fitness evaluation, e.g. by introducing dose-volume histograms (DVHs) for the target and critical regions~\cite{Xing1999,Hristov2002,Michalsky2004,Yang2004} instead of using classical least-squares.
Using different sets of importance factors applied to the target volume and critical structure objectives with a least-square error function leads to many optimal solutions (Pareto front)~\cite{Cotrutz2001,Lahanas2003,Schreibmann2004}. In most cases, the optimization of the fluence distribution is based on analytical gradient techniques.

The gradient methods 
are often trapped in local optima, contrary to 
meta-heuristic approaches,
among which genetic algorithms are frequently used in IMRT to optimize specific parameters: the effectiveness of genetic algorithms has been proven for optimizing the angles, shape and weight of the beams. Only one or two parameters are usually optimized by the genetic process
whereas other parameters are fixed by the medical physicist or determined with a fast gradient-based method~\cite{Li2004,Lei2008,Chaomin2008,Cotrutz2003,Li2003}.

\subsection{Optimization of proton therapy treatments with a genetic algorithm}
\label{ssec:hga}

In this paper, we present an inverse planning model for the spot scanning technique, in which all parameters are left free and are included in the optimization procedure. We assume that spots can be generated from any source point in space and can reach any position in the target volume (the necessary beam energy values are adjusted accordingly). Although this is technologically unrealistic, this scenario is interesting to investigate innovative irradiation patterns for existing or future beam delivery systems. Moreover it may be used to quantify the potential gain (in comparison with current treatment plans) that could be achieved by overcoming current technological limitations.


We choose to use a meta-heuristic (probability-based) optimization algorithm combined to a gradient-based method. Genetic algorithms are inspired by the concept of population evolution in biology~\cite{HandbookGA,IntroGA}. Another stochastic optimization approach is the simulated annealing, a well-spread algorithm inspired by thermodynamics and material science~\cite{Hartmann2008,Webb2005}. Compared to genetic algorithms, simulated annealing presents more robust convergence properties. However, we choose the GA approach because (i) it is very versatile and flexible in its implementation and it can be applied to a large variety of optimization problems; (ii) it becomes particularly effective when a high number of interdependent parameters (possibly of different kinds, e.g. vectors and integer values) have to be optimized, as in the case of spot scanning, where thousands of beam parameters have to be determined.

\begin{figure}[h!]
\centerline{\includegraphics[width=13.0cm]{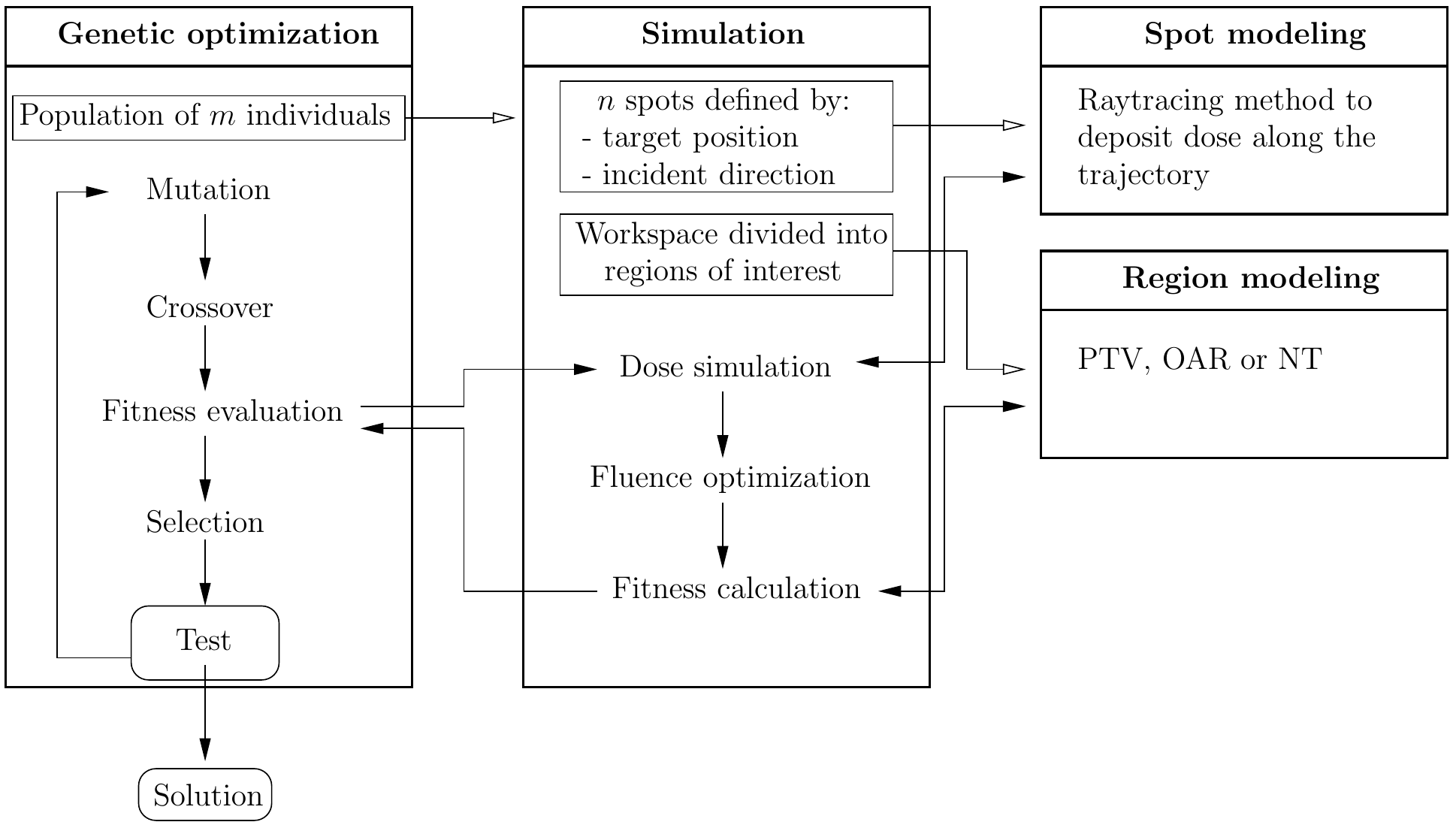}}
\caption[]{
Flow chart illustrating the structure of the proposed GA for proton beam treatment planning.}
\label{Fig_structure_HGA}
\end{figure}

The global scheme of the proposed method is presented in figure~\ref{Fig_structure_HGA}. We propose to optimize simultaneously the spot position (target point) and the direction of incidence of every beam with a GA, and the beam fluence values with a gradient descent.
The optimization loop (left box), which includes the genetic operators (mutation, crossover and selection) as well as the evaluation of the fitness, is applied to a population of $m$ individuals. Each individual corresponds to the set of parameters of a whole treatment plan. The core simulation (middle box) consists in (i) calculating the 3D dose distribution deposited by a set of beams, (ii) adjusting the fluence values of the beams (see Section~\ref{sect_fluence_optimization}) and (iii) evaluating the fitness of the considered individual. The previous steps in turn rely on the spot modeling (right box), which describes the elementary dose deposited by a single beam, and the region modeling, which defines the planning target volume (PTV), the organs at risk (OAR) and normal tissues (NT), with their corresponding dose prescriptions.



\section{Dose simulation}
\label{sect_dose_simulation}
As the GA approach requires many iterations (typically $10^3$), a fast dose simulation scheme is needed. Ray casting appears as the most suitable trade-off between speed and accuracy. The various steps and specific implementation issues of the dose calculation scheme, based on existing techniques, are summarized in the following subsections. It is worthy of note that the transportation code described below includes approximations compared to state-of-the-art Monte Carlo techniques. Nevertheless, such a model is a satisfactory trade-off (in terms of speed and accuracy) to establish the validity of the optimization algorithm.

\subsection{Beam modeling}

\begin{figure}[!h]
\centerline{\includegraphics[width=13.0cm]{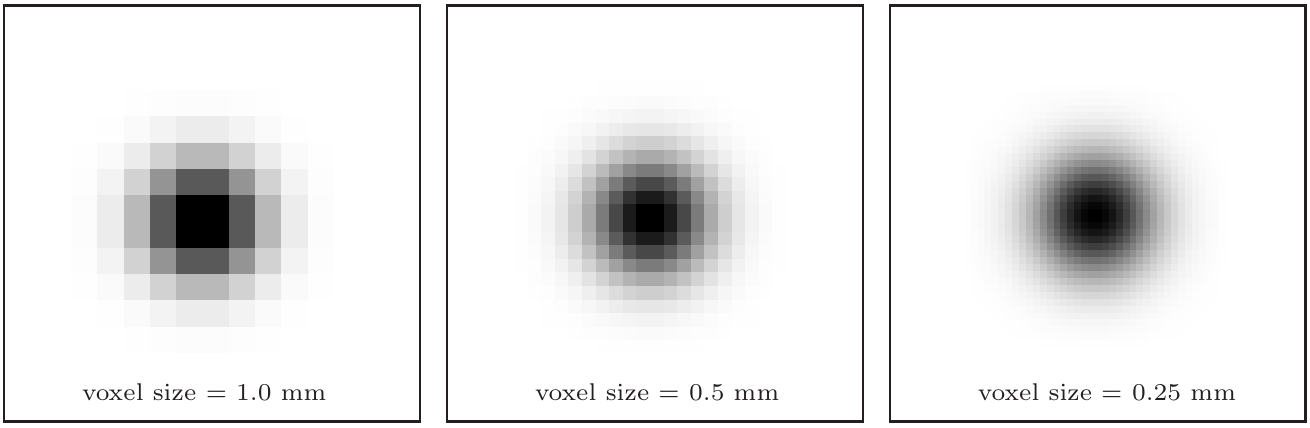}}
\caption[]{Relative fluence distribution of a Gaussian spot for different voxel sizes. In the three cases, the FWHM is $3.5$ mm.}
\label{Fig_beamSampling}
\end{figure}

We consider beams with a circular Gaussian transversal profile and negligible divergence. The beam parameters are the standard deviation $\sigma$ of the Gaussian profile (set for all beams), the target position (corresponding to the Bragg peak position) and the incident direction (described by two angles). The beam energy is not a free parameter as its value obviously determines the Bragg peak location. The beams are described using beamlets corresponding to a regular sampling in the transverse plane~\cite{Soukup2005}. The sampling interval is one half of the smallest voxel dimension and the beam support is limited to $\pm 2 \sigma$. Every beamlet is assigned a relative fluence value, as a function of its radial position with respect to the beam center.
Figure~\ref{Fig_beamSampling} presents a Gaussian spot 
sampled according to different voxel sizes.

\subsection{Energy profiles}
\label{sect_energy_profiles}

\begin{figure}[!h]
\centerline{\includegraphics[width=9.0cm]{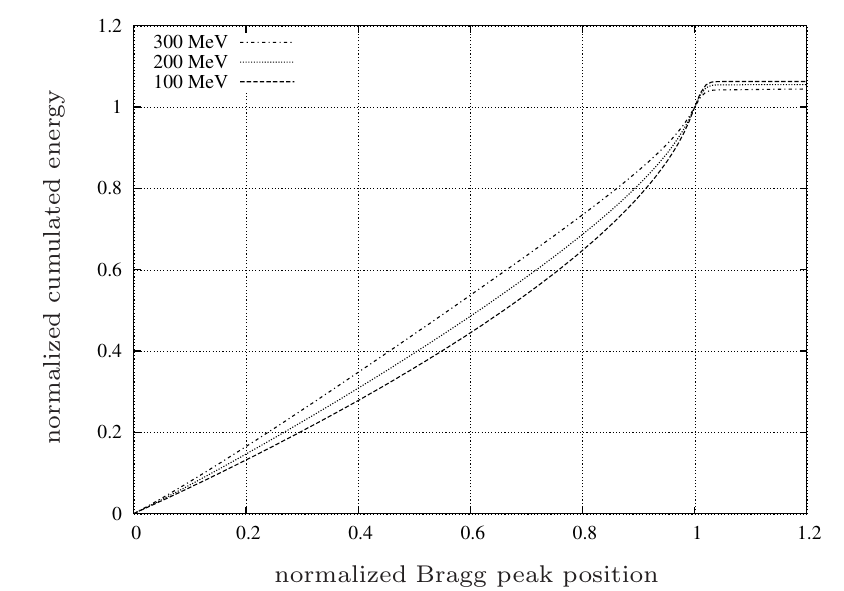}}
\caption[]{Normalized cumulative energy profiles of protons with incident energy between $100$~MeV and $200$~MeV in water. The normalized depth is unity at the Bragg peak position.}
\label{Fig_energyProfiles}
\end{figure}

Our dose deposition model is based on one-dimensional depth-dose curves expressed in terms of tabulated profiles of the cumulative deposited energy. These profiles are normalized so that the energy cumulated at the Bragg peak is unity. A set of proton cumulative energy profiles between $25$ MeV and $300$ MeV with steps of $25$ MeV has been simulated using the Geant4 Monte Carlo code~\cite{Allison2006}. Normalized profiles are shown in figure~\ref{Fig_energyProfiles} and normalization factors are listed in table~\ref{Tab_normalisationFactors}. The Bragg peak distance and corresponding cumulated energy can be obtained in a straightforward way at any incident energy using linear interpolation. The normalized cumulative energy profile can similarly be determined using linear interpolation between cumulated energy values. 

\begin{table}[h]
\begin{center}
\begin{tabular}{@{}ccc}
\hline
{$E_i$ (MeV)} & {Bragg peak (mm)} & {$E_{c}$ (MeV)} \\
\hline
$100$ & $~75.3$ & $~91.7$ \\
$150$ & $154.0$ & $136.0$ \\
$200$ & $253.4$ & $178.8$ \\
\hline
\end{tabular}
\end{center}
\caption{Bragg peak distance and corresponding cumulated energy ($E_{c}$) deposited in water per proton as a function of the incident proton energy ($E_i$).}
\label{Tab_normalisationFactors}
\end{table}

\subsection{Dose calculation based on ray casting}
\label{sect_dose_calculation}
For the sake of clarity, only physical dose will be considered here and the modeling of biological effects, which is usually not included in proton treatment planning, will be left aside.
The dose $D_{voxel}$ deposited by a beamlet in a given voxel can be calculated using the following equation:
\begin{equation}
\label{Eq_dose_voxel}
D_{voxel} = I_{beam} F_{beamlet} \frac{E_{c}(E_i,r_{out}) - E_{c}(E_i,r_{in})}{\rho_{voxel} V_{voxel}}
\end{equation}
where $I_{beam}$ is the number of incident protons in the beam, $F_{beamlet}$ is the relative fluence of the beamlet, $E_{c}(E_i,r_{in})$ (resp. $E_{c}(E_i,r_{out})$) is the cumulated deposited energy at incident energy $E_i$ at the voxel entry (resp. outlet) point, $\rho_{voxel}$ stands for the mass density of the voxel material and $V_{voxel}$ is the voxel volume.
The transport of every beamlet is described using a ray casting recipe optimized for navigation in voxelized geometry. The positions $r_{in}$ and $r_{out}$ are obtained for every voxel crossed by the ray in an incremental way~\cite{Siddon1985,Jacobs1998,Zhao2003,Freud2008}.


\subsection{Tissue inhomogeneities}

To take into account tissue inhomogeneities, we use the water equivalent path length (WEPL) method. This method consists in applying a scaling factor $k_{mat}$ to the distance $z_{mat}$ traveled in a material to obtain the corresponding WEPL as follows:
\begin{equation}
z_{W\!EPL} = z_{mat}\, k_{mat}.
\end{equation}
For a given beam, the total WEPL to the Bragg peak determines the corresponding required incident energy using table~\ref{Tab_normalisationFactors}.
The dose is still computed using equation~(\ref{Eq_dose_voxel}), in which $r_{in}$ and $r_{out}$ are now referring to WEPL distances. In our work, the scaling factors $k_{mat}$ are calculated using Geant4 simulations of the Bragg peak position in water and in the materials of interest (note that the scaling factor can be considered as independent of the incident ion energy~\cite{Batin2008}).

\section{Optimization algorithm}
\label{sect_genetic_algorithm}
The proposed algorithm combines a GA approach to optimize simultaneously the spot position (target point) and the direction of incidence of every beam with an exact-step gradient descent to determine the beam fluence values.


\subsection{Optimization of the irradiation ballistics: genetic algorithm}

In the GA approach, a population of solutions (i.e.~a finite set of individuals) evolves simultaneously and explores the landscape, i.e.~the space of all possible solutions of the problem. The set of individuals is modified at each generation by the mutation and crossover operators (search force) and by selection operators based on the fitness (selection force). A crucial point in the GA approach is the coding scheme of the individuals since these operators are directly related to it. They also have to be carefully balanced to converge towards the optimal solution. The schemata theory~\cite{Goldberg1989} and the asymptotic theory of genetic algorithms~\cite{Cerf1995} describe the convergence properties of genetic algorithms. However, this theoretical background is difficult to apply to practical problems. 

\subsubsection{Mutation and crossover operators}
\label{sect_classical_mutation_crossover}

The search of new solutions is carried out by mutation and crossover operators.
Parameters are in general represented by a series of bits. The number of bits for a parameter determines the number of discrete values in the search space, i.e.~the precision. For example, a three parameter solution with eight bits per parameter can be written as follows:

\medskip
\centerline{\includegraphics[width=6.0cm]{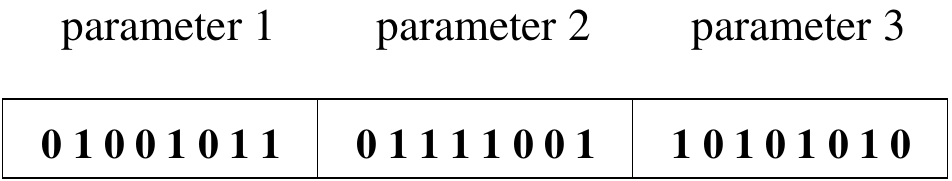}}
\medskip

In this type of coding, the search operators are rather straightforward. Mutation consists in changing a single bit value with a probability $p_m$ (generally small) or in imposing the change to a given number of bits (both methods are equivalent for a large number of mutations), which gives for example:

\medskip
\centerline{\includegraphics[width=6.0cm]{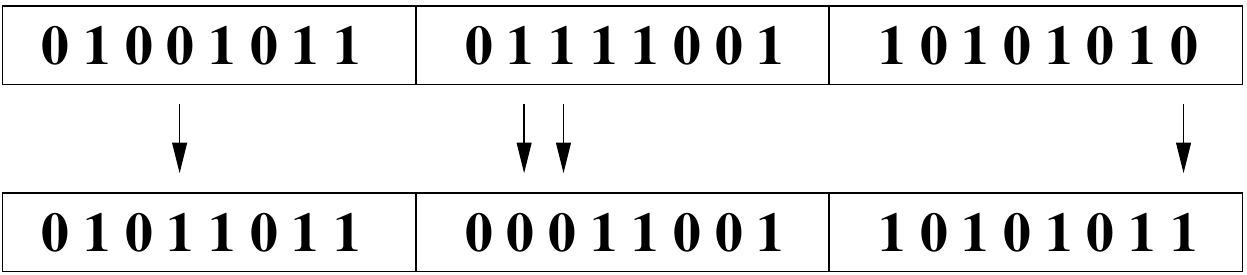}}
\medskip
Crossover consists in exchanging a part of the genetic material between two individuals with the probability $p_c$ (generally large). A cut point is randomly selected in the bit chains and the two parts are substituted:

\medskip
\centerline{\includegraphics[width=13.0cm]{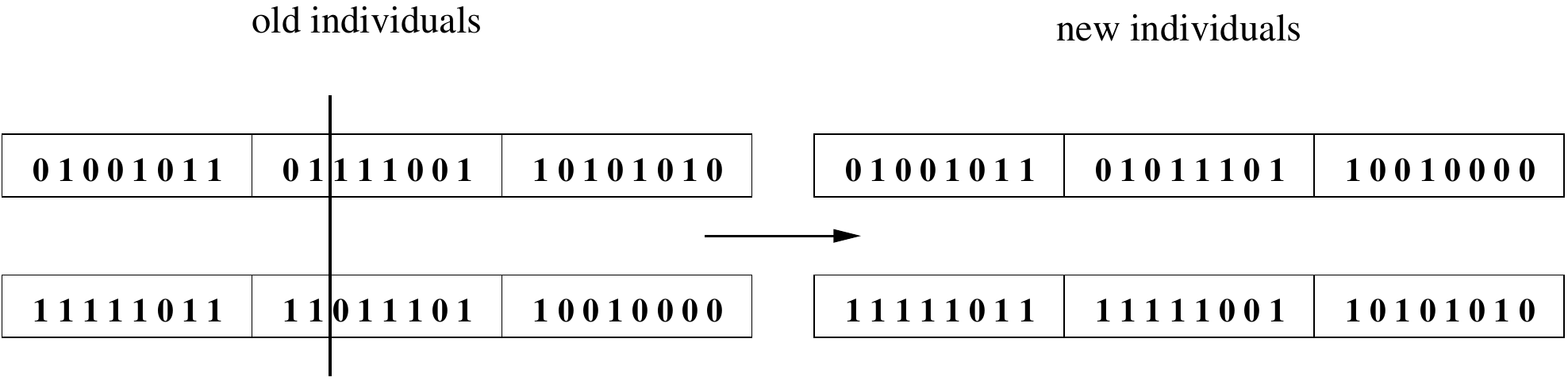}}
\medskip

The use of crossover greatly contributes to the algorithm effectiveness since it carries a fast mixing of the solutions. In general, both operators bear an efficient search tool provided mutation and crossover probabilities are correctly set. It is also important to notice that genetic operators work with a finite search space since a bit chain can only take a limited number of values.

\subsubsection{Real parameter coding}
\label{ssec:rpc}

The use of bit coding is common in the GA approach but has some drawbacks. It is easy to convert real numbers in bit chains but this can have unpredictable search consequences.
The change between two close parameter values can sometimes be very unlikely (Hamming cliffs phenomenon~\cite{Rothlauf2002}). It is worthy of note that this phenomenon is present for all parameters: the bit coding of $2^k$ is quite distant (in terms of mutations) from $2^k-1$. This is also the case for circular coding issues.

A solution to this problem is the gray coding which sees that two neighboring values always differ by only one bit~\cite{Rothlauf2002}. Another solution consists in coding individuals using a chain of parameter values, instead of a chain of bits. The genetic operators then apply directly to the parameters. This prevents differentiation between the coding of the individuals and the parameter search space. This technique is sometimes referred to as \lq real parameter coding\rq~\cite{Cotrutz2003,Li2004,Lei2008}. 
The mutation now consists in selecting a mutant parameter with a probability $p_m$ and in replacing the old value by a new one given by a mutation rule. Various mutation rules have been proposed. The most common one is the addition of a Gaussian noise to the parameter value (the standard deviation for each parameter must be chosen with care).

The parameters to be optimized in our case are the beam geometries, i.e.~a set of incidence angles and target positions (i.e.~Bragg peak positions).
We adopted the real parameter coding for our GA scheme 
with a mutation rule expressed in terms of the distance as follows:
\begin{equation}
\label{mut_rule}
P(v_1,v_2) \propto \frac {1} {d(v1,v2)}
\end{equation}
where $P(v_1,v_2)$ is the probability of mutating from $v_1$ to $v_2$ and $d(v1,v2)$ is the distance between $v_1$ and $v_2$.
This distance is a simple absolute difference in the case of incidence angles and a Euclidean norm in the case of target positions. 
It is however worthy of note that the mutation rule may become a computational burden in the case of continuous parameters with complex topology (in our case, target positions inside the tumor volume). We therefore decided to use discrete sets of parameter values, which makes it possible to pre-calculate mutation probability maps, i.e.~square matrices, the dimension of which is the number of available parameter values. The $(i,j)$ matrix element is $P(v_i,v_j)$. The azimuth beamlet angle ($\phi$, see figure~\ref{Fig_humanModel}) is a circular parameter, with value between $0$ and $2\pi$. The distance between two angle values is always taken in the $[0;\pi]$ interval.
The mutation rule (\ref{mut_rule}), simple as it may seem, makes it possible to carry out a local search on parameters without ignoring distant values. The crossover operator remains unchanged, with a cut point randomly selected in the parameter chains of two individuals.

\subsubsection{Selection operator}

The purpose of the selection operator is to assign to each individual its reproduction probability. During this stage, it is quite common to see a fit individual being selected more than once and some poorer ones disposed of. The selection law must not be too strong because good (but not optimal) individuals may come out from combinations of bad ones but the selection law must also favor the best individuals in order to propagate their attributes among the population. This operator is usually based on the score given by the fitness function: a highly adapted individual will have a high fitness value and therefore a high probability of being selected.
However, such selection probabilities directly based on fitness values may lead to a premature convergence where a very good individual screens all the others.

\begin{figure}[!h]
\centerline{\includegraphics[width=9.0cm]{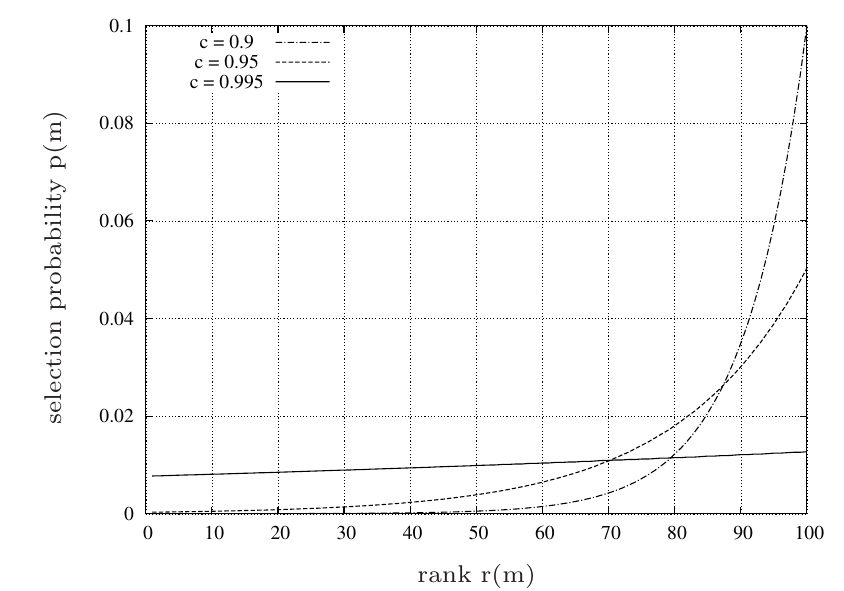}}
\caption[]{Exponential ranking selection law for a population of $M = 100$ individuals and different values of parameter $c$. The selection law tends to be uniform when $c$ is close to 1 and becomes more discriminating with decreasing values of $c$.}
\label{Fig_expRanking}
\end{figure}

To deal with this issue, the calculation of the selection probability of an individual can simply be expressed in terms of its rank instead of its fitness score, as in the exponential ranking selection scheme~\cite{Blickle1996}. Note, as a consequence, that a ranking scheme is sufficient (e.g.~based on an objective function, see subsection~\ref{sect_fitnessFunction}) and the explicit definition of a fitness function is not required. This method prevents premature convergence and conveniently fosters the mixing of individuals with different scores. The selection probability is then:
\begin{equation}
p(m) = \frac{c-1}{c^M-1} c^{(M-r(m))}
\end{equation}
where $r(m)$ is the rank of individual $m$ and $c$ is a parameter affecting the selection force. Note that the best individual is ranked $M$, the number of individuals. This method presents the advantage of having fixed selection probability values along the optimization process. In figure~\ref{Fig_expRanking}, selection probabilities are represented for different values of $c$, with a population of $M = 100$ individuals. As the parameter $c$ decreases, the best individuals tend to increase their probability of surviving. We found that a value of $c = 0.95$ is a good compromise when $M=100$. 
The corresponding selection probability curve (see figure~\ref{Fig_expRanking}) is scaled so that it can be used for any size of population.

\subsubsection{Objective function}
\label{sect_fitnessFunction}

The purpose of the objective function is to assess the adaptation of an individual to its environment. It makes it possible to rank every individual of the population. The choice of the objective function reflects in a quantitative manner the priorities given to PTV, OAR and NT, which results from medical expertise. In treatment planning, the environment is typically represented by a dose map which consists of the dose prescription to the PTV and the dose upper limits to OAR and NT. The objective function compares this prescribed dose map to a dose map given by a set of parameters (the individual). A very common type of objective function gives a score using squared residuals restricted to a set of regions of interest~\cite{Schreibmann2004,Li2004} as follows:
\begin{equation}
\label{equ_fitnessFunction}
f_{obj} = f_{PTV} + f_{OAR} + f_{NT}
\end{equation}
with
\begin{equation}
\label{equ_quadraticFitnessFunction}
f_i = \frac{1}{N_i^{\ast}} \sum_{n=1}^{N_i} \delta_{i,n} (d_n - p_i)^2
\end{equation}
where $f_{obj}$ is the total score of the objective function, $f_i$ is the region score, $n$ is the $n^{th}$ voxel in the $i^{th}$ region containing $N_i$ voxels, $d_n$ is the deposited dose, and $p_i$ is the dose prescription. $\delta_{i,n}$ is a boolean constraint flag, which is set to 1 except for OAR and NT if $d_n \le p_i$. Thus the aim of $f_{PTV}$ is to get to the prescribed dose values whereas the dose just has to be lower than a limit in the case of $f_{OAR}$ and $f_{NT}$.

Note that the value of the objective function for each region is normalized. The normalization factor $N_i^{\ast}$ is set to the number of voxels in the region for PTV and OAR. As the NT region often has many more voxels than other regions, $N_{NT}^{\ast}$ represents the average number of NT voxels crossed be the beams of an individual at the initialization step.
If the objective functions were not normalized in each region, one mutation on the incident direction of a beam would lead to much larger fluctuations in NT than in OAR or PTV since the beam crosses on average more NT voxels. This normalization of the objective function makes it possible to balance the fluctuations between regions and thus to effectively optimize several objectives simultaneously.

Another important remark is the fact that the expression of $f_{NT}$ given above tolerates the irradiation of a large volume of tissues at dose levels slightly exceeding the limit $p_{NT}$ set by the prescription. Another form, such as
\begin{equation}
\label{equ_linearFitnessFunction}
f_{NT} = \left( \frac{1}{N_{NT}^{\ast}} \sum_{n=1}^{N_{NT}} \delta_{{NT},n} |d_n - p_{NT}| \right)^2
\end{equation}
would favor the minimization of the integral dose to NT, i.e.~the choice of shorter paths to the tumor. This discussion will be illustrated in a simple example in subsection~\ref{subsection_C}.

\subsubsection{Population size and bootstrap technique}

A good balance between search and selection forces may not be sufficient to achieve an effective GA optimization. The population size $M$ plays a crucial role 
because it determines the quantity of genetic material taking part in the evolution process. Since the search is based on the mixing of different individuals, the exploration 
capacity is high when the population is large and the search more efficient.
As the population size increases, fewer generations are required to explore the same quantity of the landscape (the exploration capacity is only driven by the product of the population size by the number of generations).
However for larger populations, the number of fit individuals likely to be found at each generation is directly related to the population size, 
and the loss of a good individual does not induce a very noticeable effect since the genetic mixing is sufficient to explore several good directions simultaneously. Therefore, the choice of the population size results from a trade-off between time 
and effectiveness of the GA optimization. In our optimization, the population size is related to the number of beams 
(the higher the number of beams, the larger the population) and typically fixed between $50$ and $150$ individuals.


The initialization method is also a serious issue. Parameters can be initialized with given values or at random. In both cases, the first generations of the optimization process are not efficient because most solutions are far from optimum. The so-called \lq bootstrap\rq\ technique has been adapted to GA to deal with this issue. It consists in initializing at random more individuals than necessary and to select the best ones. In our optimization, we initialize 10 times more individuals than the initial population $M$ and the $M$ best ones are then selected to be the first generation of the GA. This method makes it possible to begin the optimization with better individuals whose genetic diversification is more productive.

\subsection{Fluence optimization: plain gradient with analytical solution}
\label{sect_fluence_optimization}

The following parameters have to be set for each beam to complete a ballistic solution: the incidence angles, the target position and the beam fluence (the beam energy is automatically determined to have the Bragg peak at the target position). It is worthy of note that the beam fluence is a very sensitive parameter in the inverse planning. The fluence of a given beam strongly depends on the fluences of other beams with which it shares voxels. Changing the fluence of one beam breaks the equilibrium in all voxels touched by this beam, which results in an erratic behavior.
We therefore propose (i) to withdraw the beam fluence parameter from the GA optimization and (ii) to transfer the optimization of the beam fluence values to a separate routine, invoked for each newly generated individual.

Most TPS in radiotherapy and ion beam therapy use gradient-based methods to optimize the fluence~\cite{Li2004}, the remaining parameters are fixed by the medical physicist. Standard gradient-based methods in TPS are the plain and conjugate gradient techniques.
Following a comparison between those algorithms in their original forms~\cite{Gemmel2008}, we have chosen plain gradient with analytical solution (PG). In order to avoid conflicts of objectives -- minimizing the dose in OAR while targeting the prescription in the PTV -- we choose to restrict the fluence optimization to the PTV whereas the OAR objective is geometrically optimized by the choice of incidence angles in the GA. To be in accordance with our GA optimization which regularly calls this fluence optimization routine, the same objective function $f_{PTV}$ of equation~(\ref{equ_fitnessFunction}) is used.

Equation~(\ref{equ_fitnessFunction}) can be rewritten with a dot product as follows:
\begin{equation}
f_{PTV}(\textbf{x}) = \sum_{n=1}^{N_{PTV}} (\textbf{a}_n \cdot \textbf{x} - p)^2
\end{equation}
where 
$\textbf{x}$ is the fluence vector and $\textbf{a}_n$ is the vector of elementary doses deposited in the $n^{th}$ voxel by each beam (the vector size is the number of beams). PG aims at minimizing $f_{PTV}$ along gradient direction $\textbf{g}^{(i)}$ for each iteration $i$:
\begin{eqnarray}
\textbf{g}^{(i)} = - \nabla f_{PTV}(\textbf{x}) = - 2 \sum_{n=1}^{N_{PTV}} \left(\textbf{a}_n \cdot \textbf{x}^{(i)} - p\right)\textbf{a}_n \nonumber \\
\textbf{x}^{(i+1)} = \textbf{x}^{(i)} + \mu \textbf{g}^{(i)} \\
\frac{\mbox{d}f_{PTV}\left(\textbf{x}^{(i+1)}\right)}{\mbox{d}\mu} = 0 \nonumber
\label{equ_step}
\end{eqnarray}
where $\mu$ is the exact step for which the objective function derivative is equal to zero. In practice, the exact step $\mu$ can be analytically calculated with the following expression:
\begin{equation}
\mu = \frac{\sum_{n=1}^{N_{PTV}} \left(p-\textbf{a}_n \cdot \textbf{x}^{(i)}\right)\left(\textbf{a}_n \cdot \textbf{g}^{(i)}\right)}{\sum_{n=1}^{N_{PTV}} \left(\textbf{a}_n \cdot \textbf{g}^{(i)}\right)^2}\mbox{.}
\end{equation}

The stopping criterion of the fluence optimization is passed when the relative difference in $f_{PTV}$ between two successive iterations is less than $0.1$ \% or when the initial value of $f_{PTV}$ has been divided by an arbitrary factor of $10^9$ (optimization is extreme). The second condition is necessary to avoid lengthening the optimization when the relative difference in $f_{PTV}$ remains constant between two iterations. Since the fluence optimization routine is called many times during the GA optimization, we took great care in implementing the information structure. Two matrices containing only the relevant information are constructed: (i) for each voxel the data needed are the beam indices and the dose they deposit per unit fluence in that voxel; (ii) for each beam, the list of voxels touched is stored.

\subsection{Specific optimization improvements}

\subsubsection{GA setting strategy}
\label{sect_GA_setting_strategy}

The balance between the selection and search forces is a crucial step in GA. We have already pointed out that it is preferable to set the selection force constant by using the exponential ranking law. Several studies also proved that the crossover operator with a one-point cut and a constant rate of $0.8$ is very stable and gives satisfactory results~\cite{Cotrutz2003,Li2003,Goldberg1989}. Setting the mutation rate is more difficult. It is the only parameter left which can balance the algorithm since the selection force and crossover rate are constant. It must be tuned carefully because its effectiveness depends on the problem to be optimized. 
When the search force is too weak (compared to the selection force), 
the genetic diversity decreases and may lead to premature convergence. Conversely, when the search is too strong, the evolution becomes purely random.
In the case of a constant mutation rate, the common way to set this parameter is by trial and error. To deal with this issue, we introduced a variable mutation rate (similar to the cooling schedule in simulated annealing). Starting with an initial mutation rate of $0.01$ --~a rather high value~-- we use the following empirical rule: if no better solution is found during $10$ iterations, the rate decreases by $10 \%$. The decreasing factor was chosen small enough to obtain a smooth mutation rate reduction.

\subsubsection{Multi-threading} Genetic optimization demands powerful computing resources. Since the mutation, crossover and selection operators are usually fast, objective function calculations constitute the bulk of the optimization workload. For each individual, the objective function requires a dose map to be calculated over the whole volume, involving heavy matrix operations. Fortunately, the GA approach is intrinsically parallel, as the same operation is repeated for each individual. In our work, we took advantage of this property using multi-threading. 
Every thread carries out the dose calculation, 
the fluence optimization with PG and the objective function calculation of one individual.
Due to the large quantity of instructions processed by the threads, the overhead (thread launch, communication, etc.)~is negligible and the actual computation time is inversely proportional to the number of processors.


\section{Behavior of the proposed optimization method}
\label{sect_results}

The behavior of the proposed optimization method was studied in two different cases: (i) an artificial C-shaped tumor volume in a water cube and (ii) a brain tumor in a CT model of a patient head. Beside the voxelized model which specifies the materials or types of tissues, 
the PTV, OAR and NT regions are given in an additional 3D label map, with a separate file specifying the dose prescription (or constraint) for each region. In the following subsections the optimization and ballistic parameters are detailed and the convergence behavior is qualitatively discussed.

\subsection{Case of a C-shaped tumor in a water phantom} \label{subsection_C}
\begin{figure}[ht]
\centerline{\includegraphics[width=5.0cm]{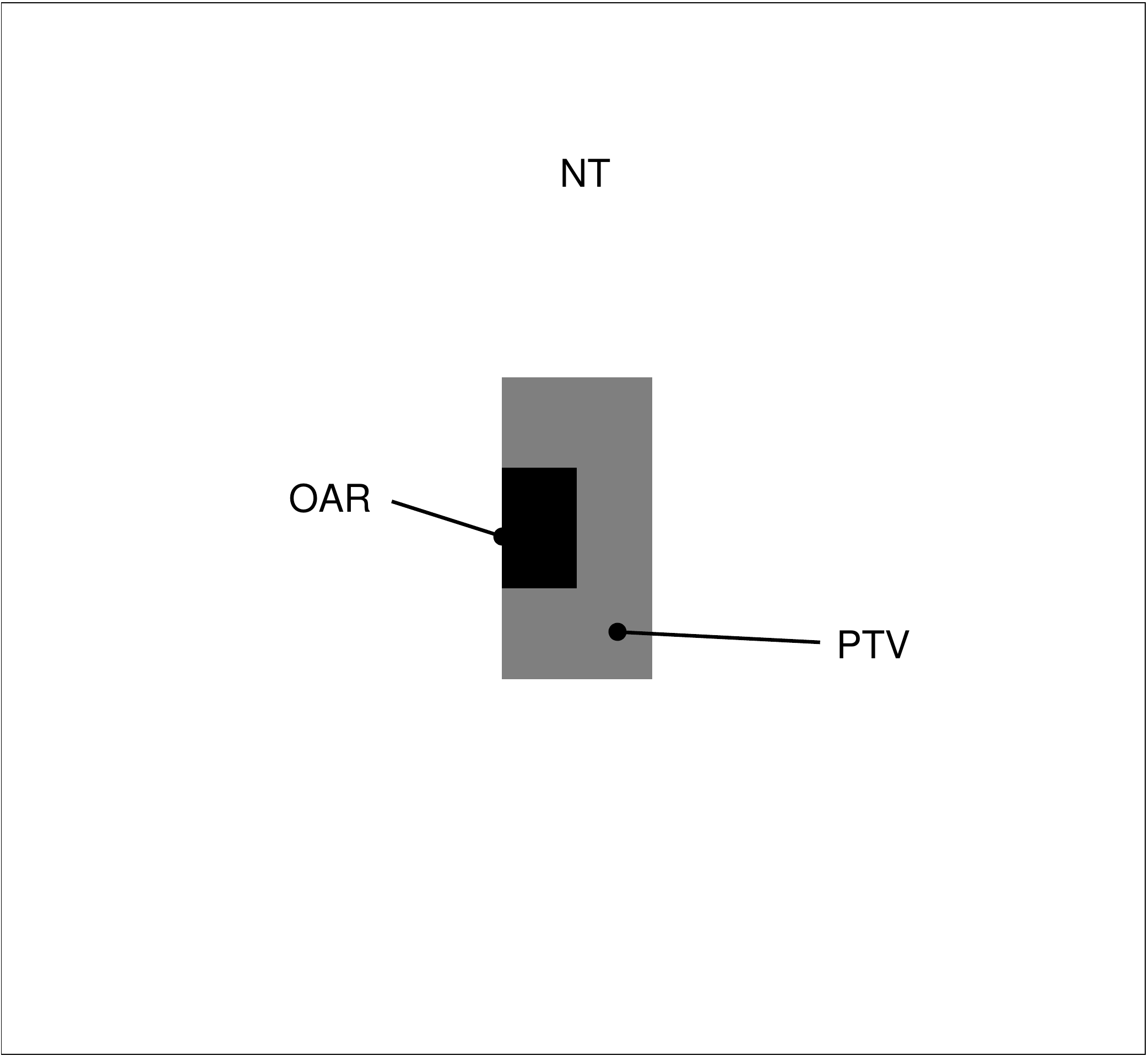}}
\caption[]{C-shaped tumor model. The phantom consists of $64 \times 64 \times 64$ cubic voxels of water, with a 1~mm voxel size.}
\label{Fig_cShapeModel}
\end{figure}

The phantom geometry is illustrated in figure~\ref{Fig_cShapeModel}. We first present some results obtained with the fluence optimization based on the PG method. Our GA optimization technique is then assessed in this simple test case. To constrain the optimization process, an OAR is placed in the C-shape. The volume of the tumor is $960$ mm$^3$. The PTV prescription is set to $1$ Gy and the OAR and NT dose limits to $0$ Gy and $0.01$ Gy respectively. The beam parameters to be optimized are the position, which is sampled at the PTV voxel centers (see discussion about the mutation rule in section~\ref{ssec:rpc}), and one angle, which is regularly sampled over $2\pi$ with $360$ values. The second beam direction angle is not optimized and is fixed to $\pi/2$ in order to have a horizontal coplanar irradiation. Despite the fact that only one angle is considered, the problem is still 3D since the tumor vertically extends over 5 slices. We irradiate the tumor with $4$ mm FWHM proton beams, described by 144 beamlets.

\subsubsection{Fluence optimization routine}
For different numbers of beams, we studied the behavior of the objective function (initial and final values) 
and of the convergence time. In this test case, target positions (voxel centers) are randomly generated in the PTV whereas beam directions are set to $0$ (see figure~\ref{Fig_cShapeDoseMaps} left). The fluences are initialized to a constant value calculated to approximately match the prescription. The results are presented in table~\ref{Tab_PGvsCG}. We observe that the final values of $f_{PTV}$ decrease when the beam number increases because of the larger number of degrees of freedom. The convergence time depends on the number of voxels in the PTV and on the number of beams that reach this region. Note that the fluence optimization is fast -- about $280$~ms for $1000$ beams -- which is appreciable since it is invoked by the GA optimization for each individual. A number of $400$ beams seems to be a good compromise between the decrease of $f_{PTV}$ and time.


\begin{table}[!h]
\begin{center}
\begin{tabular}{@{}cccc}
\hline
beams & $f^{initial}_{PTV}$ (Gy$^2$) & $f^{final}_{PTV}$ (Gy$^2$) & time (ms) \\
\hline
$~100$ & $5.39 \times 10^{-1}$ & $~1.60 \times 10^{-1}$ & $~3.5$ \\
$~200$ & $5.28 \times 10^{-1}$ & $~6.86 \times 10^{-2}$ & $~7.4$ \\
$~300$ & $5.28 \times 10^{-1}$ & $~3.78 \times 10^{-2}$ & $~14.7$ \\
$~400$ & $5.23 \times 10^{-1}$ & $~2.34 \times 10^{-2}$ & $~23.3$ \\
$~500$ & $5.19 \times 10^{-1}$ & $1.66 \times 10^{-2}$ & $~37.6$ \\
$1000$ & $5.17 \times 10^{-1}$ & $3.39 \times 10^{-3}$ & $277.0$ \\
\hline
\end{tabular}
\end{center}

\caption{Fluence optimization using PG for different beam numbers. $f_{PTV}^{initial}$ and $f_{PTV}^{final}$ are respectively the initial and final values of $f_{PTV}$.}
\label{Tab_PGvsCG}
\end{table}

\subsubsection{GA optimization}
\begin{figure}[!h]
\centerline{\includegraphics[width=11.0cm]{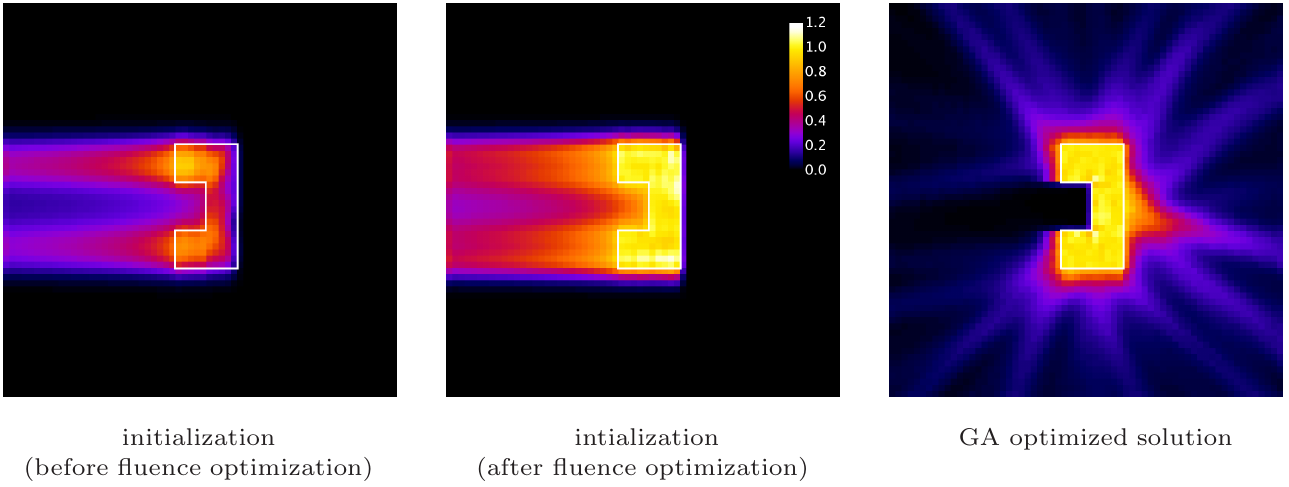}}
\caption[]{GA optimization results for the C-shaped model. The initial best dose maps are represented before (top, left) and after (top, right) the fluence optimization stage.
We optimized the initial dose map using a quadratic NT objective function (bottom, left) and a linear NT objective function (bottom, right).
Both final dose maps are almost homogeneous in the PTV and spares the OAR. The quadratic NT objective is improved by the reduction of high dose levels whereas the linear NT objective favors the shorter paths to the PTV.}
\label{Fig_cShapeDoseMaps}
\end{figure}

The number of beams is now set to $400$. The target position of each beam is randomly initialized but the angle is set to $0$, as if there was only one irradiation field in the initial state (see figure~\ref{Fig_cShapeDoseMaps}). It is worthy of note that the chosen initialization was intentionally set to the worst one-field ballistics of irradiation as the OAR region is fully irradiated and a NT region receives a high dose. The GA population is fixed to $50$ individuals. All refinements described in section~\ref{sect_genetic_algorithm} are used except for the bootstrap technique which is not relevant in the present case because every individual of the first iteration has almost the same genetic material (initialization of the angle with a constant value).
The stopping criterion chosen for the optimization process is the mutation rate value: when the product of the mutation rate by the total number of optimized parameters (number of beams multiplied by the number of beam parameters) falls below $1$ mutation per individual, the optimization is stopped. In this case, the stopping criterion is satisfied when the mutation rate falls below $(400 \times 2)^{-1}$.

The initial dose map is represented in figure~\ref{Fig_cShapeDoseMaps} (top). In the first dose map (figure~\ref{Fig_cShapeDoseMaps}, left), which represents the best individual of the first generation, the beam fluences have not been optimized and are set to a constant value. The second dose map (figure~\ref{Fig_cShapeDoseMaps}, right) is obtained after the fluence optimization stage. We can see that PG is able to find a correct set of fluences and partially achieves the homogeneity objective of the PTV region.
We optimized this initial dose map with two different NT objective functions: the quadratic function defined in equation~\ref{equ_quadraticFitnessFunction} and the linear function defined in equation~\ref{equ_linearFitnessFunction}. Besides the choice of the NT objective function, previously described genetic attributes are the same. Both GA optimized dose maps are represented in figure~\ref{Fig_cShapeDoseMaps} (bottom).
In both optimizations, the dose in PTV is almost homogeneous and the OAR is spared. As expected, the main difference lies in the NT region. We see that the quadratic NT objective case is improved by the reduction of high NT dose levels leading to a larger number of irradiated NT voxels. At contrary the linear NT objective favors the shorter paths to the PTV so as to minimize NT integral dose, increasing the dose in some trajectories. Note that the dose in NT region is not taken into account in current planning techniques. In clinical TPS the dose in NT region directly depends on the proton field entry ports which are predetermined. In our optimization framework the introduction of a specific NT objective function makes it possible to release this constraint on the number of entry ports. The choice of either strategy for NT region falls to the physician.

\begin{figure}[!h]
\centerline{\includegraphics[width=9.0cm]{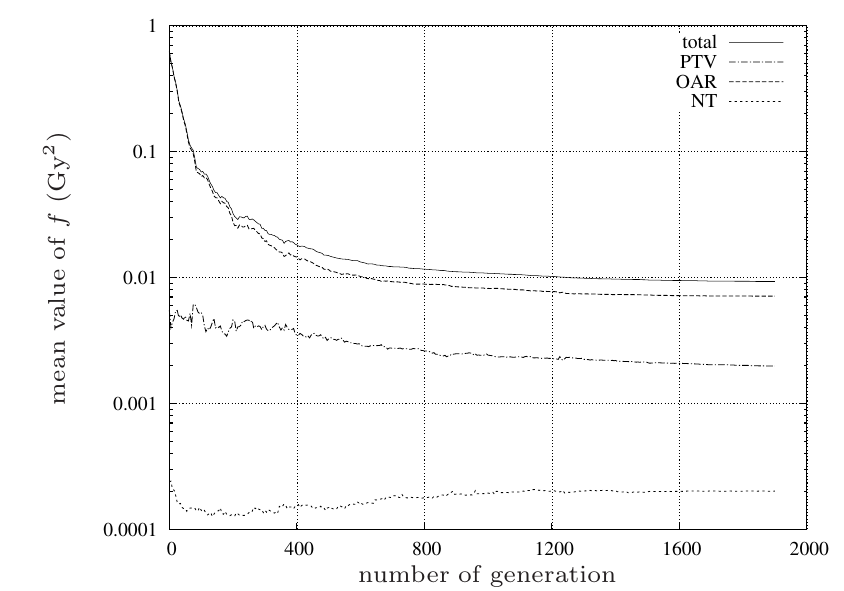}}
\caption[]{Convergence curves of the mean value of the objective function $f$ for the C-shape model optimization using quadratic (top) and linear (bottom) NT objective. For both optimization, the three contributions ($f_{PTV}$, $f_{OAR}$ and $f_{NT}$) are represented.}
\label{Fig_cShapeConvergence}
\end{figure}

Figure~\ref{Fig_cShapeConvergence} shows that the contributions to the total objective function are quite different: at the beginning $f_{OAR}$ represents more than $90$\% and $f_{PTV}$ less than $0.4$\%. The bad value of $f_{OAR}$ is due to the single irradiation field in the initial state and the good value of $f_{PTV}$ to the gradient step  (see figure~\ref{Fig_cShapeDoseMaps}, top). The convergence of a GA optimization process is essentially driven by the fluctuations induced by mutations. Note that the fluctuations of the mean value of $f$ are more pronounced when the mutation rate is high. This is clearly visible for the first $400$ iterations. In both optimizations, all objectives are optimized simultaneously but not at the same speed. The OAR objective is far from optimum and is easier to improve contrarily to the PTV objective which remains constant (top) or is slightly improved (bottom). The same distinction can be made between NT objective optimizations. Due to initial state, the quadratic NT objective (top) has a greater margin of improvement than the linear one (bottom).


\subsection{Case of a brain tumor in a CT scan}

\begin{figure}[!h]
\centerline{\includegraphics[width=13.0cm]{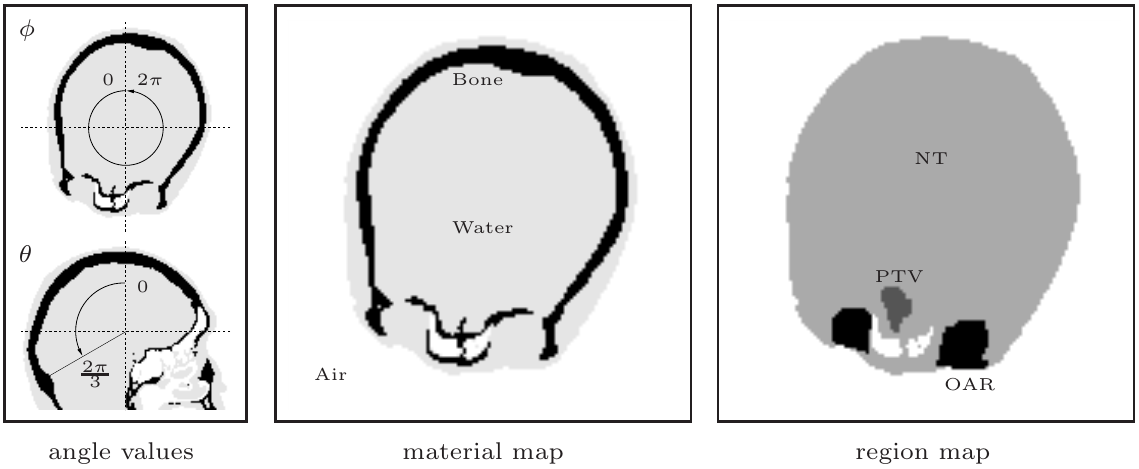}}
\caption[]{Human head model derived from a CT scan. The beam angle intervals are represented on the left. The calculation volume is an array of $128 \times 128 \times 96$ voxels with a corresponding voxel size of $1.633 \times 1.633 \times 1.5$ mm$^3$. The $46^{th}$ slice of the volume is represented as a segmented CT map (air, water and bone) in the middle and as a region map (PTV, OAR and NT) on the right.}
\label{Fig_humanModel}
\end{figure}

We now propose to investigate a clinically more relevant case based on a CT model of a human head (see figure~\ref{Fig_humanModel}). The tumor volume is located in the vicinity of the eyes, which are considered as OAR. The rest of the volume is labeled NT. The PTV prescription is set to $1$ Gy and the OAR and NT dose limits to $0$ Gy and $0.5$ Gy respectively (see equation~\ref{equ_fitnessFunction}).

The FWHM of the proton beams is set to $8$ mm (typical values in intensity modulated particle therapy lie between 5 and 10 mm). The number of beams is usually fixed by considering that (i) the effective range of dose deposition around the Bragg peak is in a crude approximation equivalent to a sphere with a diameter equal to a third of the beam FWHM and (ii) all beams altogether spread over the whole PTV~\cite{Kramer2000}. Then the following equation may be used to calculate the number of beams: 
\begin{equation}
N_{beams} = \frac{V_{PTV}}{\frac{4}{3} \pi (\frac{FWHM}{6})^3}
\end{equation}
where $V_{PTV}$ corresponds to the volume of the PTV region. According to this rule, we chose to use $1260$ beams with $8$~mm FWHM, which is sufficient to irradiate a tumor volume of $12.5$~cm$^3$. In the GA optimization, the beam positions are still sampled at the PTV voxel centers ($3125$ values), the angle $\phi$ (see figure~\ref{Fig_humanModel}) in the transverse plane is sampled from $0^\circ$ to $359^\circ$ (360 values) and the angle $\theta$ to the superior-inferior axis is sampled from $0^\circ$ to $119^\circ$ (120 values). Positions and angles are randomly initialized. The population size is set to $100$ individuals. In this optimization, all improvements described in section~\ref{sect_genetic_algorithm} are used, including the bootstrap technique. The stopping criterion chosen for the optimization process is unchanged, i.e.~when the mutation rate falls below $(1260 \times 3)^{-1}$.

\begin{figure}[!h]
\centerline{\includegraphics[width=10.0cm]{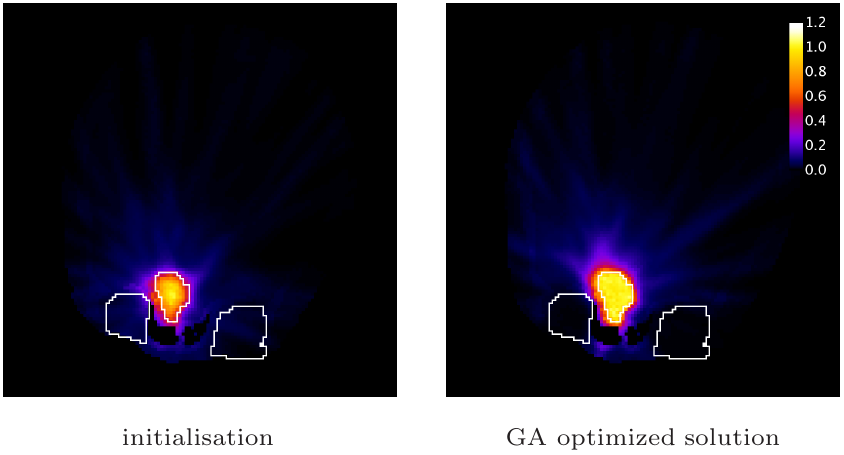}}
\caption[]{Optimization results for the brain tumor case. The dose map ($46^{th}$ slice) of the best individual is shown at the initialization stage after the PG procedure (left) and after optimization (right).}
\label{Fig_humanDoseMaps}
\end{figure}

\begin{figure}[!t]
\centerline{\includegraphics[width=9.0cm]{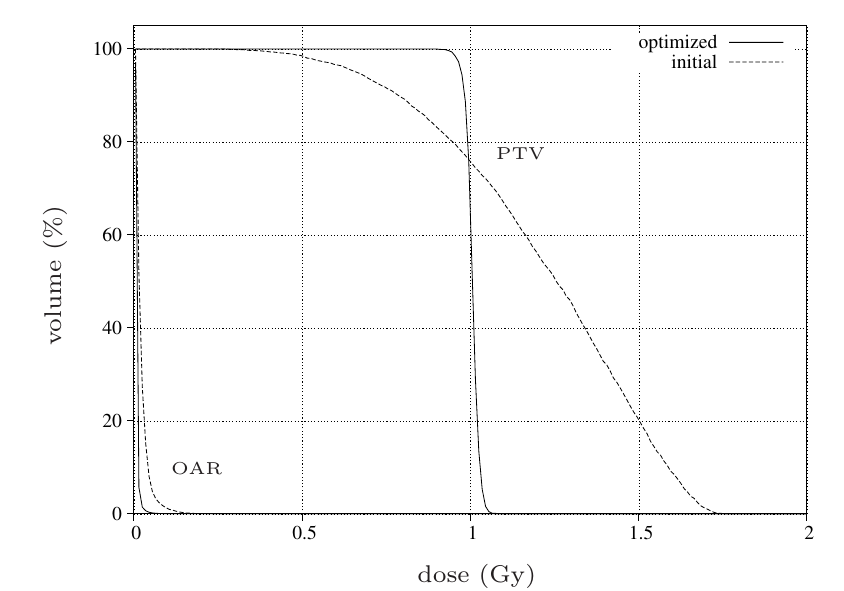}}
\caption[]{{DVHs} for the brain tumor case. The PTV and OAR {DVHs} of the best individual (after fluence optimization) are shown at the initialization (dashed lines) and after optimization (solid lines). The NT DVHs are not represented here because no major difference is visible between the initial and the optimized DVHs for this region.}
\label{Fig_humanDVH}
\end{figure}

\begin{figure}[!h]
\centerline{\includegraphics[width=9.0cm]{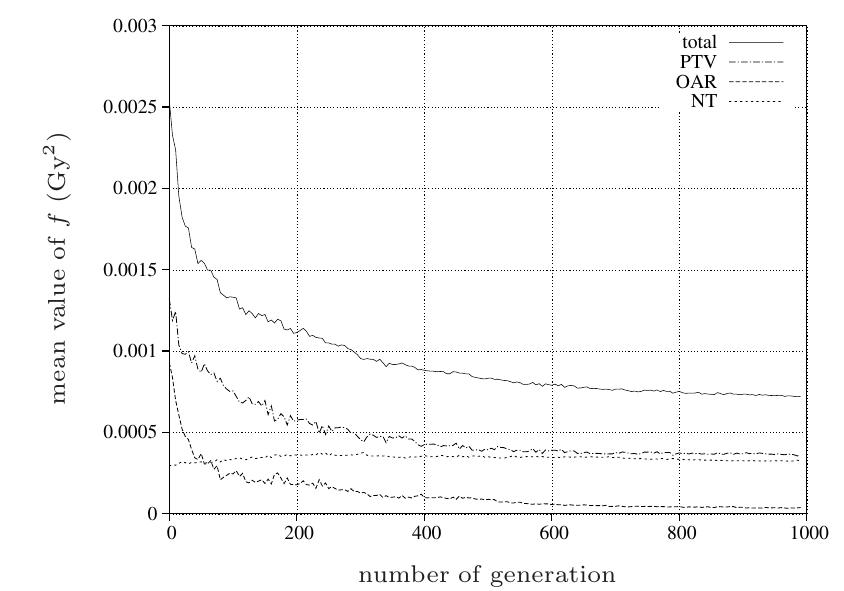}}
\caption[]{Mean value of the objective function in the case of the brain tumor.}
\label{Fig_humanConvergence}
\end{figure}

Figure~\ref{Fig_humanDoseMaps} (dose maps) and figure~\ref{Fig_humanDVH} (DVHs) show that the GA optimized solution fulfills the PTV objective and spares the OAR region. No major differences are visible between initialized and optimized dose maps for NT. The random initialization is close to be the best choice for NT because the quadratic form of the objective function (equation~\ref{equ_fitnessFunction}) prevents the dose from accumulating in the same voxels. The shortest path to the PTV cannot be always selected and as a result the beam tracks spread out during the optimization. The evolution of the mean value of the objective function is presented in figure~\ref{Fig_humanConvergence}.

The initial contributions are very different compared to the previous model. The OAR objective is almost satisfied from the start because of the bootstrap step and of the small OAR extent in the search space. The NT contribution is large in spite of the objective function normalization by the number of voxels. Our GA scheme is adequately balanced since all objectives benefit from the optimization. It is worthy of note that the beam width determines the effectiveness of the convergence of the PTV objective. It would be better satisfied with a higher number of thin beams.


\section{Concluding remarks and perspectives}

\label{sect_conclusion}

In this paper, we presented a new inverse planning technique intended to explore new schemes of treatment delivery with future enhanced technologies for spot scanning active delivery mode. Our optimization framework is based on a GA approach combined with a gradient-based method. It accounts for multiple degrees of freedom such as beam fluences, target points and irradiation directions at the same time without any {\it a priori} technological constraint. Our algorithm is versatile in the choice of both the number of objectives to be optimized and their analytical forms.
Several issues such as the real parameter coding, an exponential ranking selection law, a variable mutation rate and a bootstrap technique have been addressed to make the optimization scheme more effective. To speed up the GA optimization, we used a fast ray casting technique for dose deposition, which was multi-threaded to make the best use of the computing resources. This method has also been implemented in a modular way:
\begin{itemize}
\item Several regions of interest with different dose prescriptions or constraints can be optimized simultaneously.
\item The initialization of the parameters to be optimized can be chosen by the user (random or particular beam layout).
\item The user still has the possibility to suppress degrees of freedom to account for any technological and/or clinical {\it a priori}.
\end{itemize}

We successfully validated our method by carrying out optimizations for a C-shaped tumor in a water phantom and for a real brain tumor in a CT scan. 

Several aspects are worth being further investigated:
\begin{itemize}
 \item The coding of continuous parameters, notably for target positions in topologically complex PTV volumes. This involves implementing additional procedures as regards mutation rules.
 \item The inclusion in the GA of a variable genome size in order to let the number of beams evolve for each individual.
 \item The benchmark of our approach against state-of-the-art evolutionary strategies (Covariance Matrix Adaptation Evolution Strategy or CMA-ES)~\cite{Hansen2003}.
\end{itemize}


\section*{Acknowledgment}
We are grateful to Dr J.-F. Adam (Universit\'e J. Fourier, INSERM U-836, ESRF, Grenoble, France) for providing the human head CT scan.



\bibliographystyle{unsrt}
\bibliography{article_HGA_arxiv.bib}







\end{document}